\documentclass[12pt]{iopart}
\usepackage{ae}
\usepackage{epsfig}
\begin{document}

\title[Thin polymer films]{Single chain structure in thin polymer films:\\ 
Corrections to Flory's and Silberberg's hypotheses\footnote{Dedicated to Lothar Sch{\"a}fer on the occasion of his 60$^{\rm th}$ birthday.}}
\author{A. Cavallo,$^1$ M. M\"uller,$^2$ J.P. Wittmer,$^3$
A. Johner,$^3$\\ and K. Binder$^1$}

\address{
(1) Institut f\"ur Physik, Johannes Gutenberg Universit\"at, 
Staudinger Weg 7, D-55099 Mainz, Germany\\
(2) Department of Physics, University of Wisconsin-Madison, 
1150 University Avenue, Madison WI 53706-1390\\
(3) Institut Charles Sadron, 6 Rue Boussingault, 67083 Strasbourg, France
}

\begin{abstract}
Conformational properties of polymer melts confined between two hard
structureless walls are investigated by Monte Carlo simulation of the
bond-fluctuation model. Parallel and perpendicular components of chain
extension, bond-bond correlation function and structure factor are computed and
compared with recent theoretical approaches attempting to go beyond Flory's and
Silberberg's hypotheses.
We demonstrate that for ultrathin films where the thickness, $H$, is smaller
than the excluded volume screening length (blob size), $\xi$, the chain size parallel to
the walls diverges logarithmically, $R^2/2N \approx b^2 + c \log(N)$ with $c
\sim 1/H$.
The corresponding bond-bond correlation function decreases like a power law,
$C(s) = d/s^{\omega}$ with $s$ being the curvilinear distance between bonds and
$\omega=1$. 
Upon increasing the film thickness, $H$, we find -- in contrast to Flory's
hypothesis -- the bulk exponent $\omega=3/2$ and, more importantly, 
a {\em decreasing} $d(H)$ that gives direct evidence for an {\em enhanced}
self-interaction of chain segments reflected at the walls.
Systematic deviations from the Kratky plateau as a function of $H$ are found 
for the single chain form factor parallel to the walls in agreement with
the {\em non-monotonous} behaviour predicted by theory.
This structure in the Kratky plateau might give rise to an erroneous estimation
of the chain extension from scattering experiments.
For large $H$ the deviations are linear with the wave vector, $q$, but are very
weak. In contrast, for ultrathin films, $H<\xi$, very strong corrections are
found (albeit logarithmic in $q$) suggesting a possible experimental
verification of our results.
\end{abstract}

\submitto{\JPCM}

\maketitle

\section{Introduction}

The excluded volume interaction or self-repulsion of the segments along an
isolated chain leads to self-avoiding walk statistics of long polymers in
dilute solution 
\cite{Flory,degennesbook,DE86,CJ90,grosberg,schaeferbook,Rubinstein}.  
The chain extension as measured by the mean squared end-to-end distance, 
$R^2$, increases like a power law $R \sim N^{ \nu_{\rm SAW}}$ with chain 
length $N$.  
The exponent adopts the non-trivial value $\nu_{\rm SAW}=0.588\cdots$.
The chains swell in order to reduce intramolecular interactions
\cite{Flory,schaeferbook}. 

In a semi-dilute solution or a melt, however, these excluded volume
interactions are screened on length scales larger than the size of
the ``blob", $\xi$ \cite{Flory,degennesbook,DE86,schaeferbook}.  
For distances smaller than $\xi$ the chain statistics obeys self-avoiding 
statistics while for larger distances the chains adopt Gaussian conformations 
which are characterised by the exponent $\nu=1/2$.  
On large distances the chains can not reduce the segmental repulsion by 
increasing their spatial extension, the majority of excluded volume 
interactions stems from intermolecular contacts \cite{Flory}.

The concept of screening of excluded volume interactions in a dense polymer 
melt -- the Flory hypothesis -- plays a pivotal role in modern theory of 
polymer melts and concentrated solutions. It has been corroborated by Edwards 
\cite{DE86,ME81} and is born out of renormalisation group calculations
\cite{schaeferbook}.
The description of chain conformations as Gaussian random walks lays
at the heart of many analytical or numerical approaches like, for instance, 
the self-consistent field theory \cite{M98,SCFT} which has been widely used to 
describe spatially inhomogeneous polymer systems like surfaces of polymer 
melts, interfaces in polymer blends or self-assembly of copolymer systems.

\begin{figure}[t]
{\centerline{\epsfysize=0.3\textwidth \epsffile{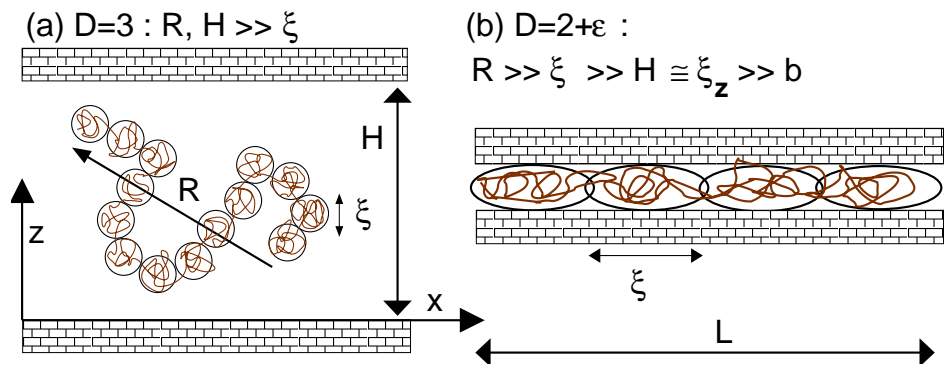}}}
\caption {Schematic illustration of a polymer chain with typical size 
$R$ confined between two parallel hard and structureless walls 
at a distance $H$. The $z$-axis is perpendicular to the lower wall.
Periodic boundary conditions are used in $x$ and $y$ direction.
(a) When $\xi \ll R \ll H$, $\xi$ being the excluded 
volume blob size, coils far from the walls are unperturbed, 
{i.e.}, $R_{\alpha}=R_{{\rm bulk}}$ for all components
$\alpha=x,y,z$.
According to Flory and Edwards, they can be represented, 
to first approximation,
as Gaussian chains of blobs of size $\xi$.
For chains (or chain sections) close to the walls it is generally
assumed that they may be reflected at the walls but remain Gaussian. 
(b) For $H\simeq \xi \ll R_{{\rm bulk}}$, 
the coils become effectively two-dimensional,
but are still allowed to overlap and to cross
in the two-dimensional projection.
These are called $D=2+\epsilon$
systems in contrast to a purely $D=2$ films where intersections 
are forbidden and the coils become compact. 
It is illustrated that in the $D=2+\epsilon$ case, 
the correlation length parallel to the film surfaces differs from $H$. 
}
\label{FIG.1} 
\end{figure}

One consequence of screening of excluded volume interactions on length scales
larger than $\xi$ is the behaviour in a thin film (see Fig.~\ref{FIG.1})\cite{SJ03}:
If the chains were describable by random walks 
without {\em long-ranged} interactions
the properties parallel and perpendicular to the film surfaces would decouple. 
According to Silberberg's hypothesis \cite{silberberg:82} the chain 
conformations could be conceived as random walks reflected at the film 
surfaces, i.e., parallel chain extensions would remain unperturbed.

As we decrease the film thickness, $H$, the chain folds back into the volume
occupied by their coil and other chains are expelled from that region.
Silberberg's hypothesis has to break down if the self-density is comparable to
the segmental number density of the bulk melt, $\rho$, i.e., $\rho R^2 H \sim
N$ \cite{M02,CMB04}.  For $H \ll \xi$ the chains adopt quasi-two-dimensional
configurations \cite{degennesbook} and the number of intermolecular contacts
are strongly reduced. This is illustrated in Fig.~\ref{FIG.1}(b).

While these phenomenological considerations yield a description of the
leading {\em scaling behaviour}
of large scale chain conformations, there are subtle
corrections that may lead to significant deviations: Renormalisation group
calculations provide great insight into the way excluded volume interactions
are screened upon increasing the density and crossing over from dilute to
semi-dilute solutions \cite{degennesbook,CJ90,schaeferbook}. 
These calculations predict the blob size, $\xi$, and the chain 
extension, $R$, as a function of the strength of the excluded
volume interaction and the segmental density, $\rho$. They also provide 
information about the crossover from self-avoiding walk behaviour to 
Gaussian statistics upon increasing the distance along the chain in 
semi-dilute solutions. 
Surprisingly, renormalisation group calculations by Sch{\"a}fer 
\cite{schaeferbook} and Monte Carlo (MC) simulations \cite{MBS00,M02}
of the bond-fluctuation model revealed deviations from the
plateau in the Kratky plot $q^2S(q)$ {\em vs.} $q$ which were traced back to a
Goldstone-type singularity. These deviations from the plateau in the Kratky
plot indicated deviations from the Gaussian behaviour on length
scales {\em larger} than $\xi$. Upon confining the chains into a thin film, one
observes an increase of the deviations from the Gaussian behaviour
\cite{M02,CMB04}.

Recently, the statistics of quasi-two-dimensional polymer melts 
(called ``self-avoiding trails") 
has been studied by Semenov and Johner \cite{SJ03}. 
This model corresponds to polymers confined to ultrathin films, 
$H \ll \xi$, where, however, 
chains and chain segments are still 
allowed to overlap as opposed to the pure two-dimensional limit 
($H\rightarrow 0$) which has been considered thoroughly in the
past resulting in beautiful analytical
predictions \cite{D2theory} and more recent numerically
tests \cite{CK90,CMB03,Y03,HP04}.
Rather than calculating the chain extension as a function of density 
and strength of the excluded volume interaction Semenov and Johner
consider the residual excluded volume interaction which
is a weak perturbation to Gaussianity in $D=3$, but is marginal,
i.e., has to be renormalised, in ultrathin layers. This allows for 
quantitative predictions in the { melt regime} both for films and
in the bulk. 
In contrast to Silberberg's argument, the chain extension for these
$D=2+\epsilon$ systems is predicted to exhibit logarithmic corrections,
\begin{equation}
R_{\alpha}^2(N)/2N = b(H)^2 + c(H) \log{N}
\label{eq:Re_slit}
\end{equation}
for the parallel components ($\alpha=x,y$) of the end-to-end distance
with respect to the walls. 
The first coefficient, $b$, corresponding to the statistical segment length
is predicted to depend only very weakly on $H$ \cite{SJ03}.
The second coefficient $c(H)$ should be positive and inversely proportional 
to the number of particles per unit surface and, hence, 
inversely proportional to the film thickness, $H$. 
Likewise, one finds in the bulk ($D=3$):
\begin{equation}
R_{\alpha}^2(N)/2N = b^2 (\rho) - c(\rho)/\sqrt{N}
\label{eq:Re_bulk}
\end{equation}
for any component $\alpha=x,y,z$ of the chain end-to-end
vector \cite{PRL04}. 
Note that the second coefficient decreases now inversely with the 
monomer density. 
In both cases, the coefficient $c$
of the leading correction term does not depend on the excluded 
volume parameter, $v$, expressing the fact that the corrections
are due to the large-scale incompressibility of the polymer melt 
\cite{SJ03,PRL04}.

In our manuscript we numerically investigate the chain statistics 
in thin polymer films paying particular attention to residual excluded 
volume effects and their dependence on the film thickness. 
As sketched in Fig.~\ref{FIG.1}, our computational study covers the 
regime from thick films, $H > R$, to ultrathin films, $H < \xi$, 
and allows us to investigate the crossover from the bulk behaviour ($D=3$) 
over ultrathin films ($D=2+\epsilon$) 
up to purely two-dimensional systems ($D=2$). 
In the next section we describe the model and simulation technique. 
Then we investigate the chain extensions parallel and perpendicular 
to the film, the bond-bond correlation function 
and the single chain structure factor.
The final section presents our conclusion.

\section{Computational model and some technicalities}

%\fbox{BFM general}\\
Coarse-grained models are very efficient to investigate the universal
properties of dense polymer systems. In these kind of models, one 
integrates over the microscopic degrees of freedom, which do not 
affect the universal physical laws but only their prefactors and 
local properties.
In this work we have used the ``bond fluctuation model" (BFM) --
a lattice Monte Carlo scheme where a monomer occupies $8$ lattice 
sites (i.e., the volume fraction $\phi = 8 \rho$, $\rho$ being the number density) 
and the bonds, $\mathbf{l}$, between adjacent monomers can vary in length and 
direction, subject only to excluded volume constraints and entanglement 
restrictions \cite{BFM,BWM04}. The BFM has been extensively studied
in the past \cite{M98,MBS00,M02,CK90,CMB03,PRL04,Paul,MWBBL03}.

Chain configurations are updated by two types of canonical moves: 
random monomer hoppings and slithering-snake (or reptation) moves \cite{BWM04}.
In the first case we consider random displacements of an effective monomer by
one lattice site in one of the 6 possible directions in the lattice. 
These moves are efficient in relaxing the local properties of the chain, such
as the bond angle or the bond length. In slithering-snake moves, a segment 
of the chain is removed from one end and added to the opposite end of the 
chain in a random direction. The latter moves relax the chain conformation 
a factor $N$ faster than the random monomer hoppings.
(See ref.~\cite{MWBBL03} for details on the exponential
increase of the relaxation time for very long chains due
to correlations in the slithering-snake motion.) 

%\fbox{Slit geometry}\\

As illustrated in Fig.~\ref{FIG.1}, we use a $L \times L \times H$ box with
hard walls at $z=0$ and $z=H+1$, i.e., the thinnest film with $H=2$ corresponds 
to a pure $D=2$ system where chains are not allowed to overlap. 
For $H=4$ this becomes possible and, therefore, this is an ideal system to 
test the predictions of ref.~\cite{SJ03}. 
We have systematically varied the film thickness between $H=2$ up to $H=84$ 
while keeping the parallel dimensions to the walls fixed at $L=256$ 
(with the exception $L(H=2)=512$). 
Periodic boundary conditions have been used in $x$ and $y$ directions.
Note that $L$ is always much larger than $R_x$, the typical chain size
component parallel to the walls.

%
%\fbox{Parameter range}\\
%
The chain length ranged between $N=16$ up to $256$.
We have concentrated in this study on dense polymer melts and 
all the data reported here are for a number density $\rho=0.5/8$. 
For this density the (bulk) excluded volume length is about 
$\xi \approx 7$ \cite{Paul}.
%

%\fbox{Samples}\\
We used the configurational bias method to create the initial configurations 
for chain lengths up to $N=64$. (The longest chain length possible numerically 
with this method decreases with $H$.) To this end a chain is grown monomer by monomer
into the system taking into account all possible bond vectors for the next monomer.
From those positions of the next monomer one that does not overlap with any other 
monomer in the system, is chosen at random. This choice introduces a bias 
which is removed by the Metropolis 
acceptance criterion once the chain is fully grown.
For larger $N$, we started with relatively
compact two-dimensional coils ($R\sim N^{1/2}$) oriented parallel to the
surfaces. In all cases we equilibrate 
our systems using a mixture of local and slithering snake moves which has been 
found to be the most efficient \cite{MWBBL03}. 
Some of our configurations have already 
been used in some recent related investigations on thin polymer films 
\cite{CMB04,CMB03}.
% 

%\fbox{Measurements}\\
Various static and dynamical properties parallel and perpendicular
to the walls have been measured on the fly and for a more detailed analysis configurations were periodically stored.
Since the $x$ and $y$ directions parallel to the walls are 
equivalent these properties are averaged together. If not specified
otherwise, properties are averaged over all chains irrespective 
of their distance to the walls.
For later reference, we note here the bulk values for the end-to-end distance 
$R \approx 49.8$ and the radius of gyration $R_g \approx 20.3$
for chains of length $N=256$. 
The three components of the mean-squared bond vector 
$l_{\alpha} \equiv \left< l_{\alpha}^2 \right>^{1/2} \approx 1.52$ 
of the bulk remain unchanged for all $H > 7$. 
Obviously, the perpendicular component must ultimately vanish and the 
parallel component increase. 
For $H=4$ we find, for example, $l_x=l_y=1.63$ and $l_z=1.2$.

\section{Results}

We discuss now in turn three intra-chain properties:
the global chain size $R_{\alpha}(N,H)$,
the bond-bond correlation function $C_{\alpha}(s)$ as a function
of curvilinear distance $s$ along the molecule's backbone and the form factor 
$S_{\alpha}(q)$.

\subsection{End-to-end distance and radius of gyration}

We start the discussion by presenting the chain size
as measured by the three components 
$R^2_{\alpha} \equiv \left< (r_{\alpha,N}-r_{\alpha,1})^2 \right>$
of the end-to-end distance $R$ and the corresponding components
of the radius of gyration $R_g$. 
Here, $r_{\alpha,n}$ denotes the $\alpha$-component of the position 
of the $n$th monomer of the chain.
%

\begin{figure}[t]
\centerline{\resizebox{13.0cm}{!}{\includegraphics*{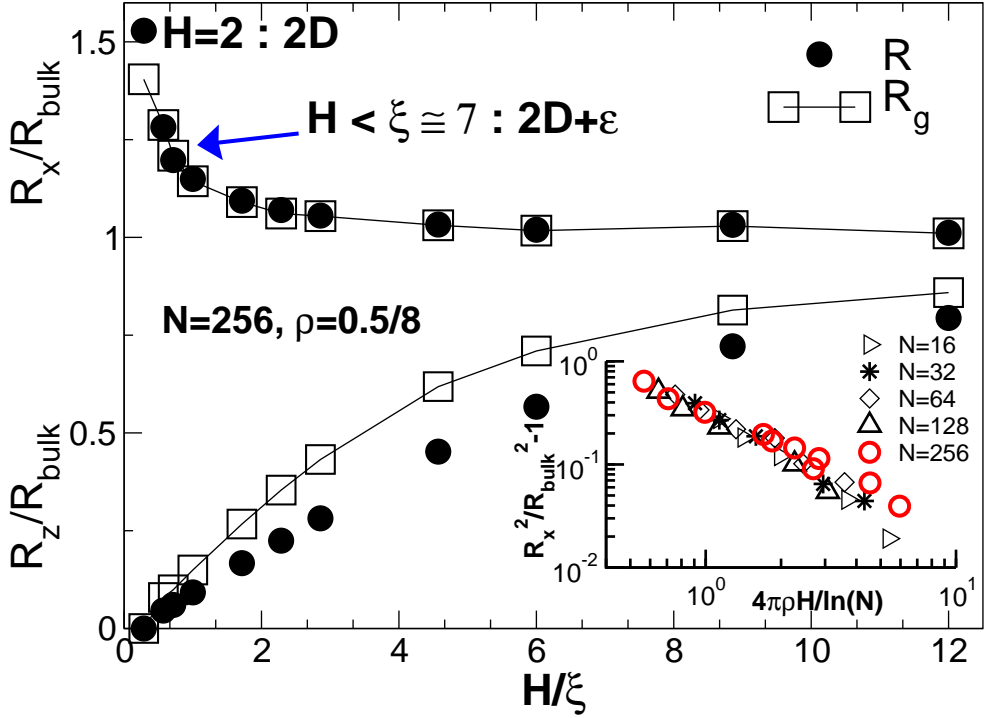}}}
\caption {Parallel and perpendicular components of
end-to-end distance $R$ (closed symbols) and
of radius of gyration $R_g$ (open symbols)
for $N=256$ as a function of the film thickness $H$. All the
quantities are normalised with respect to the corresponding bulk value. 
The data are averaged over all chains in the film irrespective of the distance 
of their centre of mass from the walls.
The (reduced) parallel components remain constant for distances
larger than the characteristic width $H^{\star}$ which for the
chain length used is of the order of $\xi$. 
The bulk value of $\xi$ for the presented
volume fraction is indicated. 
Note that for $H=2$ no chain overlap is possible in our model
and the chains are rigorously two-dimensional. $D=2+\epsilon$
behaviour is expected for $2 < H < \xi$.
The perpendicular component increases continuously with $H$.
The effect is linear for $H \ll R_{\rm bulk}$, as one expects.
The inset presents the chain length dependence of the parallel
chain dimensions as a function of $H/H^\star$ for different chain lengths, $N$,
as indicated in the key.
\label{FIG.2}}
\end{figure}

%
%\fbox{Chain size as a function of $H$}\\
In Fig.\ref{FIG.2} we present the end-to-end distance and the radius of gyration, $R_{g}$, as a function of the film thickness $H$. 
The data are normalised by the corresponding bulk value, i.e., all the ratios must become 
unity for large $H$. Only data for chain length $N=256$ has been included. 
As expected, the thinner the film, the larger is the component parallel to the 
surface and the smaller is the perpendicular one.
The perpendicular component increases continuously with $H$.  
The effect is linear for $H \ll R_{\rm bulk}$, as one expects even for 
perfectly Gaussian chains. Note that the perpendicular component of 
$R_g$ increases more rapidly as the component of the end-to-end vector for small $H$.
In agreement with \cite{SJ03} the (reduced) parallel components remain 
constant for $H \gg H^{\star} \approx \xi \approx 7$, but increase rapidly 
for thinner films. 
(We remind that for $H=2$ no chain overlap is possible in our model and 
the chains are rigorously two-dimensional.)  
This demonstrates that perpendicular and parallel components
of the polymer couple in contrast to Silberberg's hypothesis. 
Note that the observed chain length dependence of the crossover
slit width $H^{\star}$ is relatively weak for the chain lengths
we have probed (cf.~Fig.~\ref{FIG.2} inset).
This is in line with the predicted {\em logarithmic} increase
$H^{\star} \approx \xi \log(N/g)$,
$g \sim \xi^2$ denoting the numbers of monomers contained in
the excluded volume blob,  
which follows readily from the condition $\rho R^2 H^\star \sim N$ utilising 
Eqs.~(\ref{eq:Re_slit}) and (\ref{eq:Re_bulk})\cite{SJ03}. 

%\fbox{Chain size vs. $N$}
%
In the next figure we present the parallel component of the end-to-end vector, 
$R_x$, as a function of $N$ for different $H$. 
Also included is the data for bulk systems ($D=3$) without walls and periodic 
boundary conditions in all directions. In agreement with Eq.~(\ref{eq:Re_bulk}) 
we find corrections to ideality even in this latter case. For details the reader 
is referred to the recent study \cite{PRL04}. As expected, the reduced data of
both $D=3$ and $D=2$ limits ($H=2$) become chain length independent for long 
enough chains. (Note that the chain size in $D=2$ is $R_x\sim N^{1/2}$
without logarithmic correction \cite{D2theory}.)
More interestingly, we find that all intermediate data
{\em diverge logarithmically} for long enough chains (thin lines). 
Extrapolating our data, one even expects that for $N \gg 2000$ and $H=4$ 
the chains become actually larger than the pure $D=2$ ones. 
%%which ultimately become compact.
%
In their work Semenov and Johner \cite{SJ03}
have predicted for the slope in log-linear coordinates [cf.~Eq.~(\ref{eq:Re_slit})]
\begin{equation}
1/c = 4\pi c_0 = 4\pi \rho H
\label{eq:c_slit}
\end{equation}
where $c_0$ is the number of particles per unit surface.
Hence, the coefficient decreases inversely with $H$.
In contrast, 
the first coefficient, $b$, of Eq.~(\ref{eq:Re_slit}) is predicted to depend
only very weakly on $H$ \cite{SJ03}. 
Using Eq.~(\ref{eq:c_slit}) and adjusting $b(H)$ we obtain rather
good agreement with the numerical data. This is the central result of this work. 
Note that $b(H)$ is essentially constant as can be seen from the convergence
of the lines for small $N$. 
We even get acceptable fits for film widths 
three times larger than $\xi$ which is consistent 
with the logarithmic increase of $H^{\star}$ with chain 
length mentioned above.
%
%Unfortunately, a possible test by plotting $R_x(H)$ for
%much longer chain lengths as in Fig.~\ref{FIG.2} is currently
%not possible.

%
\begin{figure}[t]
\centerline{\resizebox{13.0cm}{!}{\includegraphics*{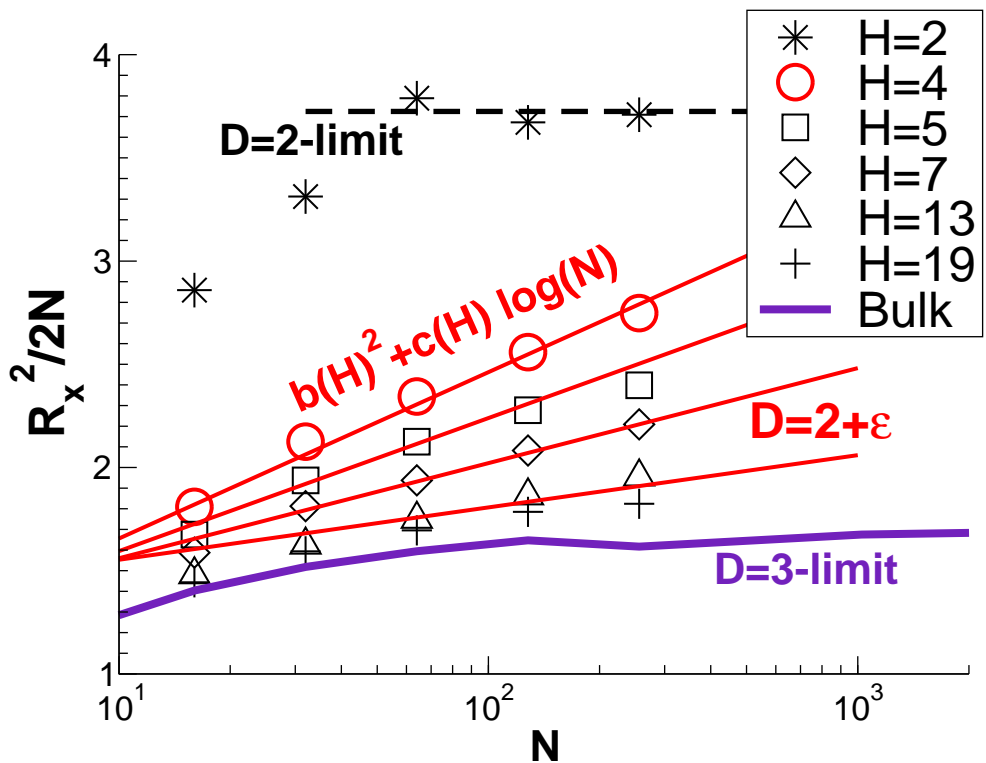}}}
\caption {Parallel components of the end-to-end distance  
for different $H$ as a function of chain length $N$.
We plot $R_{x}^2/2N$ in log-linear coordinates to demonstrate the 
expected logarithmic divergence for $D=2+\epsilon$ systems \cite{SJ03}. 
The  thin lines compare with the prediction,
Eqs.~(\ref{eq:Re_slit},\ref{eq:c_slit}), where we
directly verify $c(H)=1/4\pi\rho H$ and fit for the coefficient $b(H)$
which is found to depend very weakly on the film width
becoming ultimately $H$ independent (left side of figure).
The pure $D=2$ data ($H=2$, stars) become chain length independent 
for chain lengths $N \ge 64$ (dashed line). Note that even the bulk data is
monotonously increasing in agreement with Eq.~(\ref{eq:Re_bulk}).
\label{FIG.3}}
\end{figure}

We have also investigated the components of the mean-squared distance
along the chain
$R_{\alpha}^2(s) \equiv \left< (r_{\alpha,n} - r_{\alpha,n+s})^2 \right>_n$
obtained by averaging over all chains and all pairs of monomers separated
by a curvilinear distance $s=|m-n|$ along a chain. 
Here, $\langle\cdots\rangle_n$ denotes the average
over all couples of bonds having curvilinear distance $s$
and over all polymers in the system.
The predictions of ref.~\cite{SJ03,PRL04}, Eq.~(\ref{eq:Re_slit})
and Eq.~(\ref{eq:Re_bulk}), generalise readily from $s=N$ to arbitrary 
curvilinear distance. This is well confirmed by our data which
looks quite similar to the one given in Fig.~\ref{FIG.3}.
Instead of discussing these data we will rather proceed by
discussing the numerically more challenging {\em curvature} of $R_{\alpha}(s)$, 
i.e., by presenting our results on the bond-bond correlation function,
which allows a much more accurate test of the predictions.

\subsection{Bond-bond-correlation function}

The connection between mean-squared distance between two segments $n$ and $m$ 
along the chain and the bond-bond correlation function is due to the exact formula
\begin{equation}
\left< l_{\alpha,n} l_{\alpha,m} \right> = - \frac{1}{2}\frac{\partial^2}{\partial m \partial n}
\left< (r_{\alpha,n}-r_{\alpha,m})^2\right>
\label{eq:Cs_Rs}
\end{equation} 
valid for $\alpha=x,y,z$.
Note that the bond vector is defined as ${\bf l}_n = {\bf r}_{n+1}-{\bf r}_n
\approx \partial {\bf r}_n/\partial n$, the latter identity becoming true
in the continuous limit.
More generally, Eq.~(\ref{eq:Cs_Rs}) may be written in terms of the 
difference operator $\Delta_n f(n) = f(n+1)-f(n)$. In that sense, it remains
valid for arbitrary curves, ${\bf r}_n$, even for non-differentiable
Brownian walks.

Averaging again over all pairs $(n,m=n+s)$ we have computed
the correlation functions 
$C_{\alpha}(s) = \left< l_{\alpha,n} l_{\alpha,n+s} \right>_n/l_{\alpha}^2$.
\footnote{We have compared this definition to
$\left< e_{\alpha,n} e_{\alpha,n+s} \right>_n$ for the normalised bond
vector ${\bf e}_n$ without finding any measurable difference.
This might be due to the weak correlation between bond length and bond angle fluctuations.
}
Neglecting chain end effects and using Eq.~(\ref{eq:Cs_Rs}) one obtains the
compact relation $2C_{\alpha}(s) l_{\alpha}^2 = \partial^2_s R_{\alpha}^2(s)$
between both curvilinear properties.

From Eqs.~(\ref{eq:Re_slit}) and (\ref{eq:Re_bulk}),
generalised to arbitrary $s$, we immediately see that
the bond-bond correlation function must become  
\begin{equation}
C_{\alpha}(s) = d s^{-\omega}
\label{eq:Cs}
\end{equation}
with a prefactor $d\sim 1/\rho$ and an exponent $\omega=3/2$ for the bulk and 
$d=c/l_{\alpha}^2$, $\omega=1$ for ultrathin films.
Noticeably, the long-distance behaviour of $C_{\alpha}(s)$ is not characterised 
by a length scale (persistence length) but the residual excluded volume interaction 
change the dominant scaling behaviour to a scale-free power law.
This is in marked contrast to Flory hypothesis where long-ranged
correlations between bonds are neglected and $C_{\alpha}(s)$
is assumed to decay exponentially (or even faster) \cite{PRL04}.
The technical advantage of the bond-bond correlation function compared
with the curvilinear distance is that it does not depend on the trivial, 
but very large Gaussian contribution which drops out after differentiating.
Hence, it allows us to focus directly on the corrections to the Gaussian behaviour
on large length scales.

\begin{figure}[t]
\centerline{\resizebox{13.0cm}{!}{\includegraphics*{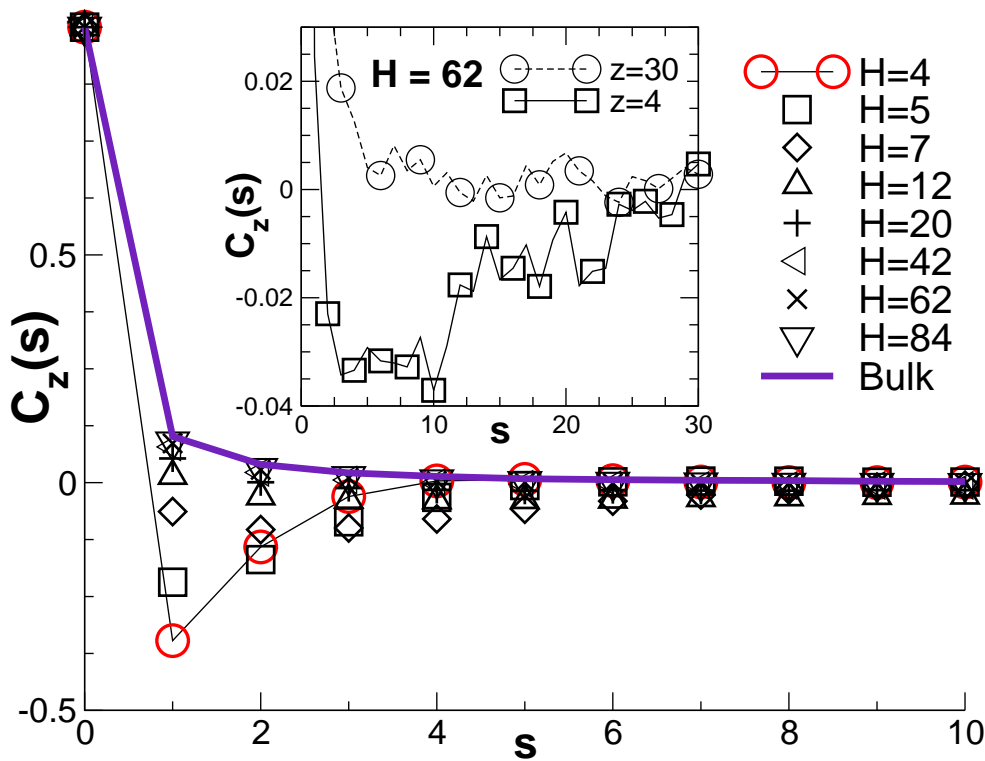}}}
\caption{Perpendicular bond-bond correlation function, $C_z(s)$, 
{\it vs.} curvilinear distance, $s$, between monomers for $N=256$ and 
different film thicknesses, $H$. The inset shows the behaviour of 
the same correlation function for polymers with centre of mass at distances 
$z = 4$ or $30$ from the wall at $z = 0$ in a film with $H = 62$.
\label{FIG.4}
}
\end{figure}

In Fig.~\ref{FIG.4} the perpendicular bond-bond correlation function,
$C_z(s)$, is plotted as a function of the curvilinear distance, $s$, for chain length $N=256$ 
and different film thicknesses, $H$.
For very thin films negative correlations at short distances can be observed.
For $H=4$ and $H=5$, on average, the angle $\vartheta$ between two consecutive bonds, 
in the direction perpendicular to the walls, is larger than $90^{o}$ 
(i.e., the two vectors have opposite direction). For
$H=7$, the correlations reach the minimum for $s=2$. This means that
chains fold back every two monomers.
These results are a trivial consequence of confinement. 
For very thin films,
due to strong fluctuations of the density as a function of the distance from the walls,
the system essentially consists in the
superposition of mono-layers with distance equal to the lattice
spacing. For $H=4$ and $H=5$ we have two layers, then two consecutive
segments along the chain are prevalently located in different planes and the chain
is reflected at the surface. For $H=7$, we
have one more layer and then up two three linked monomers can be placed
at different height. Upon increasing the film thickness, $H$, above the blob size, $\xi$,
$C_z(s)$ gradually tends to the curve obtained in the bulk.

The same effect can also be observed for thicker films taking into
account only chains with centre of mass having a distance, $z$, from the walls smaller than $\xi$.
This is a consequence of excluded volume effects in the direction perpendicular 
to the wall which give rise to a strong anisotropic deformation of the coil.
In the inset of Fig. \ref{FIG.4} we show $C_z(s)$ calculated for polymers at
different distance, $z$, of their center of mass from the surface. In this plot we
compare the curves obtained for $z=4$ and $z=30$ in a film with
$H=62$. While in the first case the interactions with the wall
must be very strong the walls should have a negligible effect on the latter
where the chains are exactly located in the middle of the film. 
Indeed, only in the first case evident negative correlations 
can be seen and we recover bulk behaviour in the second.
The insufficient statistics does not allow a precise location of the 
minimum for $z=4$. This value roughly corresponds to $s=5$ which is
close to the distance of the center of mass from the wall.

\begin{figure}[t]
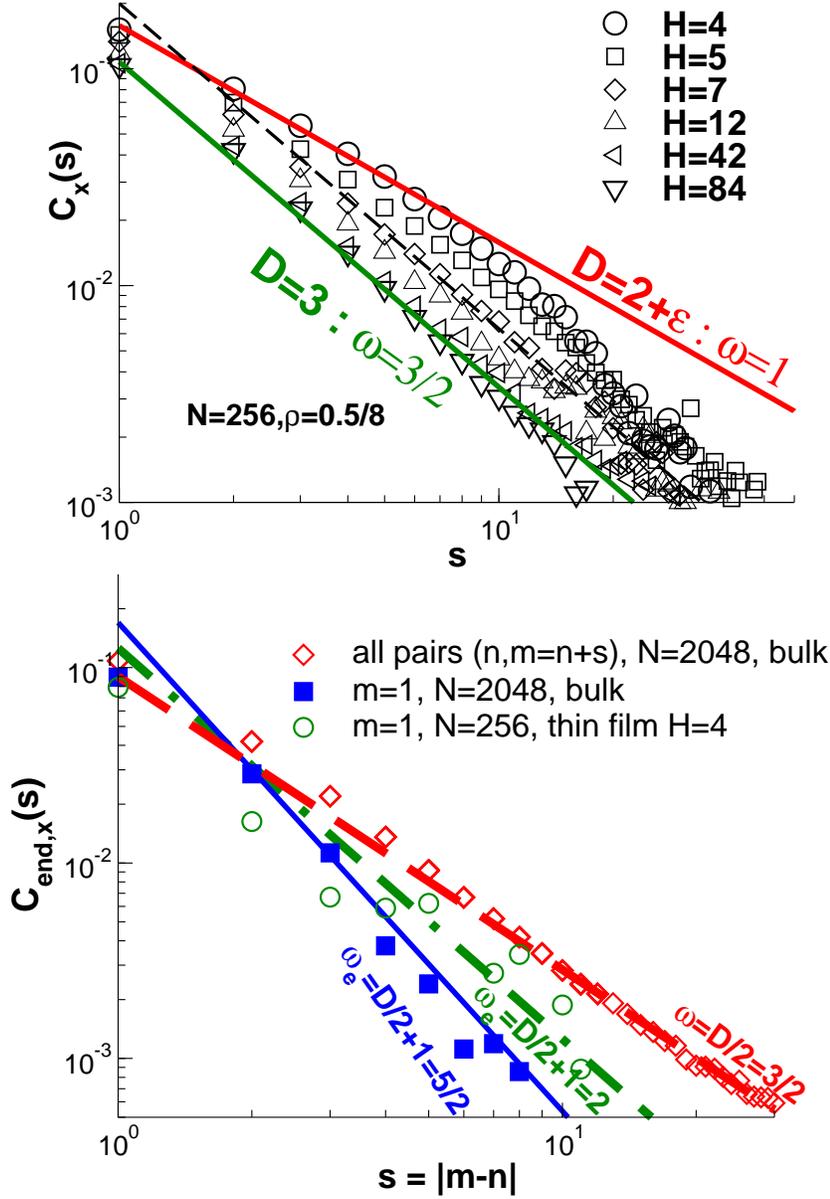

\centerline{\resizebox{10.0cm}{!}{\includegraphics*{fig5.eps}}}
\centerline{\resizebox{11.0cm}{!}{\includegraphics*{fig5b.eps}}}
\caption{(a) Log-log plot of $C_x(s)$ {\it vs.} $s$ for different film 
thicknesses and for $N = 256$. 
The lines indicate the power-law behaviour, $C_x(s) = d(H)/s^{\omega}$,
predicted for different regimes. 
The slope $\omega = 3/2$ (bottom line) is expected for bulk systems and larger slits.
While this exponent remains unchanged for $H \gg \xi$ the amplitude $d(H)$ does.
As indicated by the broken line, we demonstrate a systematic {\em increase}
with decreasing $H$.  Finally, for $D=2+\epsilon$ films our data 
confirms the exponent $\omega=1$ (upper bold line) implicit to the work
of Semenov and Johner \cite{SJ03}.
(b) Double logarithmic plot of $C_{{\rm end},x}(s)$ in the bulk 
and a thin film, $H=4$, corresponding to $D=2+\epsilon$. The bulk
correlation function averaged over all pairs is shown as reference.
The predicted power laws are indicated by straight lines.
\label{FIG.5}
}
\end{figure}

A quantitative analysis of the decay of $C_x(s)$ is given in Fig.~\ref{FIG.5}  
where we plot the function on a double logarithmic scale. The curves can be 
fitted rather well by power laws, in agreement with Eq.~(\ref{eq:Cs}),
and exponential behaviour can definitely be ruled out for long enough chains.
Obviously, this is a clear-cut contradiction to Flory's and Silberberg's
descriptions. 
For large thicknesses, $H > 7\approx \xi$, we observe the expected bulk
exponent $\omega = 3/2$. Interestingly, the prefactor, $d$, increases as we
decrease the film thickness, $H$, although the exponent remains constant. This
shift can be related to the dependence on the fraction of polymers $\propto
1/H$ which interact with the surface. Chain segments close to the wall have
increased bond-bond correlations $\left< l_{x,n} l_{x,n+s} \right>$ due to the
enhanced self-interaction of the segment which is, in turn, due to the
reflection of the segment at the wall. To first order, the reflections may
still be described by Gaussian statistics.  As $H$ increases and the relative
population of chain segments of size $s$ decreases the amplitude must decrease.
This interpretation has been directly confirmed by computing, as in
Fig.~\ref{FIG.4}, $C_x(s)$ for various distances, $z$, of the chain center of
mass from the walls (not shown).  For ultrathin $D=2+\epsilon$ films, the bulk
exponent must break down. Indeed, our data are still compatible with a
power-law but with the $\omega=1$ exponent predicted for this regime.

Since finite-chain size effects could be crucial here larger chains are needed
to rigorously establish the exponent over more than a decade in $s$.  In the
bulk the correlation function decays more rapidly for short chains and its
explicit dependence on $N$ has been obtained \cite{PRL04}.  This is, in part,
due to the behaviour close to the chain ends.  In the analysis we have averaged
the correlations over all pairs with the same distance, $s$, along the chain's
contour but neglected chain end effects. This is permissible for large
$N$.\footnote{The dependence of correlations on the position along the chain
has recently been studied for isolated, self-repelling polymer chains by
Sch\"afer and Elsner\protect\cite{L2}} Interestingly, the correlation function
measured from a chain end, $C_{{\rm end},\alpha}(s) \equiv \left< l_{\alpha,0}
l_{\alpha,s} \right>_n/l_{\alpha}^2$ decays with a different, stronger
exponent, $\omega_e=D/2+1$.  The simulation results are presented in panel (b)
of Fig.~\ref{FIG.5}. For bulk melts, very long chains are available
\cite{PRL04} and confirm the value of the exponent, $\omega_e=5/2$ in $D=3$.
The data for thin films are also compatible with the theoretical prediction.

\subsection{Form factor $S(q)$}

The single chain form factor
\begin{equation}
S(q)= \frac{1}{N} 
\sum_{n,m=1}^N
\left< \exp(i \sum_{\alpha} q_{\alpha} (r_{\alpha,n}-r_{\alpha,m}) ) \right>
\label{eq:Sqdef}
\end{equation}
of a chain is an important, experimentally relevant quantity.
The components of the scattered radiation vector are
denoted by $q_{\alpha}$.
We are only interested here at the internal correlations 
parallel to the wall and, hence, consider only wave vectors parallel 
to the surface ($q_z=0$) and omit the index in the following.

\begin{figure}[t]
\centerline{\resizebox{13.0cm}{!}{\includegraphics*{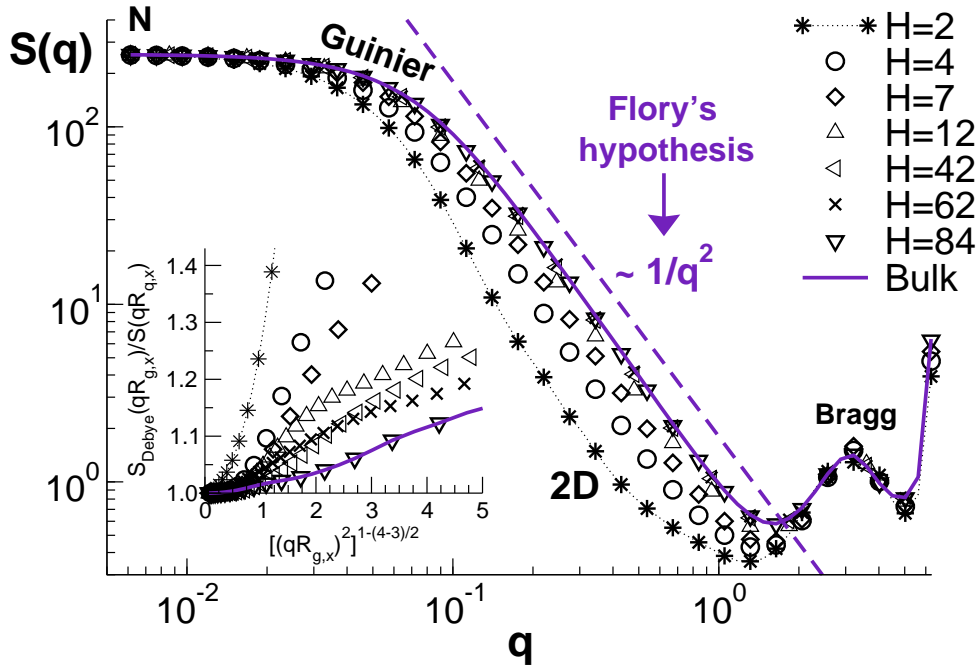}}}
\caption{Single chain form factor for wave vectors $q$ parallel
to the wall surfaces for different $H$. 
Only data for $N=256$ and $\rho=0.5/8$ are considered.
The bulk result ($H\rightarrow \infty$) is given by the bold line,
Flory's hypothesis indicated by the broken line.
The Bragg limit ($q > 1$) is not affected by the presence 
of the walls. However, this can be clearly seen for larger length scales where 
the scattering amplitude goes systematically down with decreasing film width.
The effect is the most prominent for $D=2$ system (stars), 
but even there the Bragg peak is not altered. 
The inset presents the ratio of the structure factor, $S_{\rm Debye}$, of a
Gaussian chain with the measured radius of gyration and the structure factor, $S(q)$,
observed in the simulations as a function of $\sqrt{(qR_{gx})^2}$.
}\label{FIG.6}
\end{figure}

In Fig. \ref{FIG.6} we present $S(q)$ for different film
thicknesses and chain length $N = 256$. The analog quantity, calculated for the
bulk system, is also depicted as a reference curve.
For small $q$ the form factor just counts the number of scattering units, 
for large $q$ it probes the structure on the scale of the monomers, 
the so-called Bragg peak. Not surprisingly, in both limits $S(q)$ 
does not depend on the presence of walls. 
For small $q$ the data should be described by the rather general 
expansion $S(q) = N [1 - (R_{gx} q)^2 + \cdots ]$ due to Guinier \cite{DE86}. 
Here, $R_{gx}(H)$ stands for the measured parallel component of the radius 
of gyration. Since $R_{gx}$ increases for small $H$ (Fig.~\ref{FIG.2}) 
the structure factor should decrease in agreement with what can be 
seen in the figure. One can check explicitly for all $H$ that Guinier's
formula applies. Hence, the deviation from the bulk limit
seen in this regime is fully described in terms of the
increased chain size.

The Gaussian self-similar structure for intermediate length scales is the
essence of Flory's hypothesis and results in a power-law behaviour of the form,
$S(q) \sim 1/q^2$, which is indicated by the broken line. Broadly, this is
confirmed by the thick films  ($H \ge \xi$). Differences become more apparent
when we plot the ratio of the structure factor of a Gaussian chain, $S_{\rm
Debye}$, with the same radius of gyration and the measured single chain
structure factor. This ratio is presented in the inset of Fig.~\ref{FIG.6}.
The clearly observable deviations are already present in the bulk \cite{MBS00}
and increase upon reducing the film thickness \cite{M02}. Significant
differences with respect to Gaussianity are observed for the
% $D=2+\epsilon$
thin films.

\begin{figure}[t]
\centerline{\resizebox{13.0cm}{!}{\includegraphics*{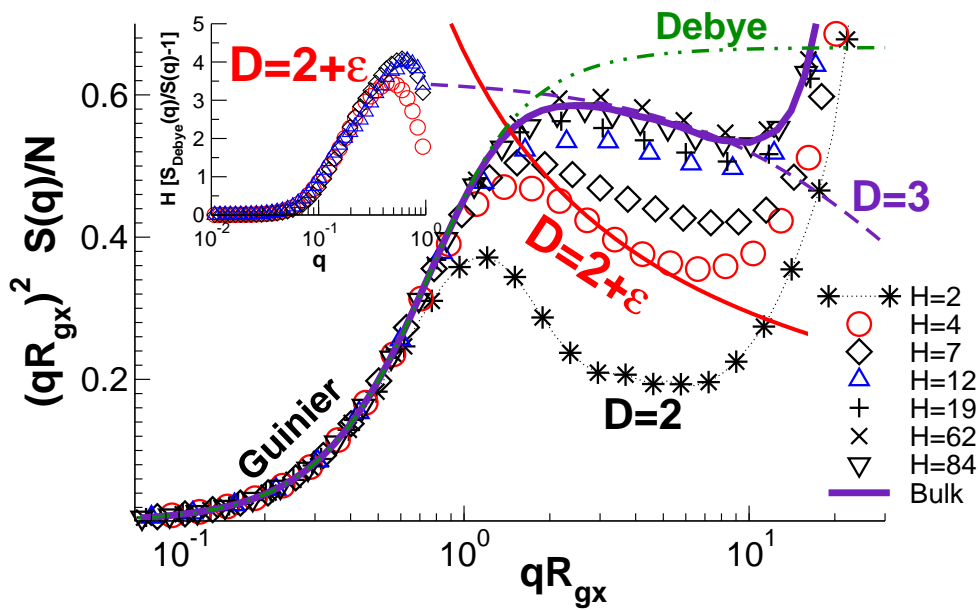}}}
\caption{ 
Kratky plot $(q R_{gx})^2 S(q)/N$ {\em vs.} $q R_{gx}$ for wave vectors 
parallel to the walls using the same symbols as in the previous plot. 
We have rescaled the data with the chain length $N$ and, $R_{gx}$,
the measured component of the radius of gyration parallel to the surface.
We see that the density fluctuations get systematically suppressed 
with decreasing film width $H$. 
The data for $H=4$ has been compared with the logarithmic correction,
Eq.~(\ref{eq:Sq_slit}), proposed in ref.~\cite{SJ03}
and the bulk data (bold line) with the corresponding
linear relationship (dashed line).
The dashed-dotted line indicates the Debye functions 
$S_{\rm Debye}(q R_{gx})/N$.
The inset presents the deviation, $S_{\rm Debye}(q)/S(q)-1$,
of the measured form factor in thin films with $4\leq H \leq 12$
from the Debye function for the same radius of gyration.
As expected, for $1 > q > 1/R_g$ the deviations grow logarithmically in $q$ and 
the amplitude is proportional to $1/H$ as can be seen from the scaling 
collapse. 
The deviation decreases again for large $q$ due to additional (non-universal) 
corrections in the Bragg limit
which are not taken into account by the theory. 
}
\label{FIG.7}
\end{figure}

The deviations from the Gaussian behaviour as well as the effect of the
confinement are magnified in the Kratky plot shown in Fig.~\ref{FIG.7} where we
have plotted $(q R_{gx})^2 S(q)/N$ {\em vs.} $q R_{gx}$ for $N=256$. This plot
illustrates a popular method for extracting the chain dimensions from scattering 
data. Using the measured value of the radius of gyration we obtain perfect 
scaling for the Guinier regime, $qR_{gx}<1$. Not surprisingly, the
scaling fails for very large $q$ (Bragg peak) where the data diverges.  More
interestingly, the scaling fails even for intermediate wave vectors where it
should hold for Gaussian statistics \cite{MBS00}. Instead we find a pronounced
{\em non-monotonous} behaviour in stark contrast to traditional hypotheses. As
already seen in the previous figure, the deviations get more pronounced with
increased confinement \cite{M02}. Similar deviations are also observed in
semi-dilute bulk solutions of long chains, $N=2048$ \cite{MBS00}.

These effects have been analytically predicted in semi-dilute solutions
\cite{MBS00} and melts \cite{SJ03}.  
For $qR_{gx} \gg 1$ and for long chains \footnote{For exponentially long chains yet another regime is
expected \cite{SJ03} which is outside the reach of simulation.}
\begin{equation}
\frac{S_{\rm Debye}(q)}{S(q)}-1 = e(H) \log (q f(H))
%R_g \quad \mbox{i.e.,} \; 
%\frac{2 N}{(qR_g)^2S(q)}=1+ f \log qR_g \quad \mbox{for} \;qR_g \gg 1
\label{eq:Sq_slit}
\end{equation}
has been suggested for thin films. Only the first of the two coefficients
$e(H)$ and $f(H)$ should depend sufficiently strongly on the slit width
to be measurable: $e(H) \sim c(H) \sim 1/H$  \cite{SJ03}.  
$S_{\rm Debye}(q)$ denotes the structure factor of a Gaussian chain with identical
radius of gyration.  
In the inset of Fig.~\ref{FIG.7} we explicitly verify the
film thickness and wave vector dependence for $4 \leq H \leq 12$.
It also confirms that $e(H)\log(f(H))$ is rather small.
For $D=3$ melts the deviation from the Kratky plot has been estimated 
more recently by Beckrich \cite{philippe} using the perturbation
approach of ref.~\cite{SJ03}. This yields the linear expression 
%\begin{equation}
$\frac{S_{\rm Debye}(q)}{S(q)} - 1 = e(\rho) q$ 
%\quad \mbox{i.e.,} \; \frac{12 N}{(qR)^2S(q)}=1+ fq \quad \mbox{for} \;qR \gg 1
%q^2 S(q) = e - f q
%\label{eq:Sq_bulk}
%\end{equation}
which we have fitted to our bulk data (dashed line). A similar expression was
derived for the limit of semi-dilute solutions ($\rho \ll 1$ but $\rho R^3/N
\gg 1$) by Sch\"afer $S_{\rm Debye}(q)/S(q)-1=e' [(q \xi)^2]^{1-\epsilon/2}+...$
where $e'$ is a constant and $\epsilon=4-D$ \cite{schaeferbook}. Thus we expect
corrections to the Kratky plot to be present over the entire concentration
regime where the chains overlap.
Since not all prefactors are explicitly available as a function of the model
parameters of the BFM the test is less rigorous than the one in
Fig.~\ref{FIG.3}. While our simulation data are compatible with the theoretical
predictions longer chain lengths would clearly be desirable.
 
\section{Conclusions}

Using Monte Carlo simulations of long polymers confined between
hard structureless walls we investigate deviations from the classical 
hypotheses of Flory \cite{Flory} and Silberberg \cite{silberberg:82}.
In agreement with recent theoretical work \cite{SJ03} on ultrathin films with 
$H < \xi$ we find that parallel and perpendicular components couple and 
the chain size parallel to the surfaces increase strongly.
We quantitatively demonstrate the logarithmic divergence, 
Eq.~(\ref{eq:Re_slit}), for the chain size parallel to the 
walls as a function of chain length, $N$. The prefactor $c \sim 1/H$ 
predicted for the logarithmic deviation allowed a rather accurate fit 
of our data. As emphasised in refs.~\cite{SJ03,PRL04}, these effects are  due
to the chain connectivity and the large-scale osmotic incompressibility of 
the solution and express universal physics independent of local properties.

Perhaps the most striking effect of this study is the power law asymptote 
for the bond-bond correlation function which measures the curvatures
of the curvilinear distance along the chains. It allows a direct numerical 
test of the non-Gaussian corrections both in the bulk and in the
ultrathin films and demonstrates the presence of long-ranged correlations
neglected by the classical hypotheses.
Note that the decay exponent for both bulk and thin film geometry
can be expressed in the form $\omega=\nu D$, $D$ being the effective 
dimensionality of the system and $\nu=1/2$ the characteristic exponent 
of a random walk. Interestingly, this is similar to the result obtained 
for the velocity-correlation function of dense liquids in two and three 
dimensions \cite{LongTimeTails}.

An important consequence of our work arises for an experimentally
relevant quantity, the static structure factor $S(q)$.
In fact, simulation and theory show distinct non-monotonous behaviour of
$q^2 S(q)$ {\em vs.}~$q$ (Kratky plot) due to the non-Gaussian corrections
which even get enhanced with decreasing film width. 
This suggests a possible route for experimental verification and 
is cause of serious concerns with regard to the standard operational 
definition and measure of the persistence length from the ``assumed'' Kratky plateau.

Finally, we point out that the physical mechanism which has been sketched 
above is rather general and should not be altered by details such as a 
finite persistence length 
--- at least not as long as nematic ordering remains negligible. 
This is in fact confirmed by preliminary and on-going simulations.

\section*{Acknowledgements}
We thank J.~Baschnagel, P.~Beckrich, S.~Obukhov and A.N.~Semenov 
for useful discussions. A.C. thanks the MPI Mainz for a fellowship. Additional 
financial support from LEA, MOLSIMU, ESF SUPERNET 
programs is acknowledged as well as computational resources by the HLR Stuttgart,
the NIC J{\"u}lich and IDRIS, Orsay.

\end{document}